\documentclass[%
reprint,
superscriptaddress,
showpacs,
amsmath,amssymb,
aps,
prl,
longbibliography
]{revtex4-1}

\usepackage{psfrag,graphicx,epsfig,color}% Include figure files
\usepackage{dcolumn}% Align table columns on decimal point
\usepackage{bm}% bold math
\usepackage{natbib}
\usepackage[usenames,dvipsnames,svgnames,table]{xcolor}
\usepackage{subfigure}
\usepackage{rotating}
\usepackage{float}
\usepackage[normalem]{ulem}

\def\re    {{R_\lambda}}

\begin{document}

\title{
Turbulence is an ineffective mixer when Schmidt numbers are large 
}

%%% Authors %%%
\author{Dhawal Buaria }
%\thanks{}
\email[]{dhawal.buaria@nyu.edu}
\affiliation{Tandon School of Engineering, New York University, New York, NY 11201, USA}

\author{Matthew P. Clay}
\affiliation{School of Aerospace Engineering, Georgia Institute of Technology,
Atlanta, GA 30332, USA}

\author{Katepalli R. Sreenivasan}
\affiliation{Tandon School of Engineering, New York University, New York, NY 11201, USA}
\affiliation{Department of Physics and the Courant Institute of Mathematical Sciences,
New York University, New York, NY 10012, USA}

\author{P. K. Yeung}
\affiliation{School of Aerospace Engineering, Georgia Institute of Technology,
Atlanta, GA 30332, USA}
\affiliation{School of Mechanical Engineering, Georgia Institute of Technology,
Atlanta, GA 30332, USA}

\date{\today}% It is always \today, today,
             %  but any date may be explicitly specified

%\thispagestyle{empty}

\begin{abstract}

We solve the advection-diffusion equation for a stochastically stationary passive scalar $\theta$, in conjunction with forced 3D Navier-Stokes equations, using direct numerical simulations in periodic domains of various sizes, the largest being $8192^3$. The Taylor-scale Reynolds number varies in the range $140-650$ and the Schmidt number $Sc \equiv \nu/D$ in the range $1-512$, where $\nu$ is the kinematic viscosity of the fluid and $D$ is the molecular diffusivity of $\theta$. Our results show that turbulence becomes an ineffective mixer when $Sc$ is large. First, the mean scalar dissipation rate $\langle \chi \rangle = 2D \langle |\nabla \theta|^2\rangle$, when suitably non-dimensionalized, decreases as $1/\log Sc$. Second, 1D cuts through the scalar field indicate increasing density of sharp fronts on larger scales, oscillating with large excursions leading to reduced mixing, and additionally 
suggesting weakening of scalar variance flux across the scales. 
The scaling exponents of the scalar structure functions in the inertial-convective range appear to saturate with respect to the moment order and the saturation exponent approaches unity as $Sc$ increases, qualitatively consistent with 1D cuts of the scalar.

\end{abstract}

\maketitle

\paragraph{Introduction:} 
A defining property of fluid turbulence,
which plays a critical role in myriad natural and 
engineering processes, 
is that it mixes substances extremely well \cite{tl72,ZW00,PD05}. 
Thus, any circumstances in which turbulence loses that property is 
naturally important  to study and understand.
This Letter examines such an instance by considering mixing 
of passive scalars with large Schmidt numbers, $Sc \equiv \nu/D$, 
where $\nu$ is the kinematic viscosity of the fluid and $D$ is 
the molecular diffusivity of the mixing substance.
By analyzing a massive database generated through state-of-the-art direct numerical simulations (DNS) of the governing equations, we show that even 
fully developed turbulence at high Reynolds number becomes an 
ineffective mixer when the $Sc$ is rendered very large.

The rate of mixing of a scalar $\theta$ is related to the average `dissipation' rate $\langle \chi \rangle$ of its variance, defined as $\langle \chi \rangle= 2D \langle |\nabla \theta|^2\rangle$. There is a general claim 
that $\langle\chi\rangle$ remains finite even when  
$D \to 0$ . This claim derives from the analogy with the mean dissipation rate of turbulent kinetic energy, which is theorized to be independent of 
viscosity when the latter is sufficiently small ($\nu \to 0$) 
\cite{taylor:1935,K41}. 
There is concrete empirical evidence that 
anomalous dissipation of kinetic energy
is essentially correct \cite{sreeni84,sreeni98,pearson02,kaneda03}.
However, whether the analogous property holds for 
scalar dissipation 
still remains an unresolved question 
\cite{batch1959a,MY.II,ShrSig00,Donzis05}.
We show that it does not when $Sc$ is large.

\begin{figure}
\begin{center}
\includegraphics[width=7.5cm]{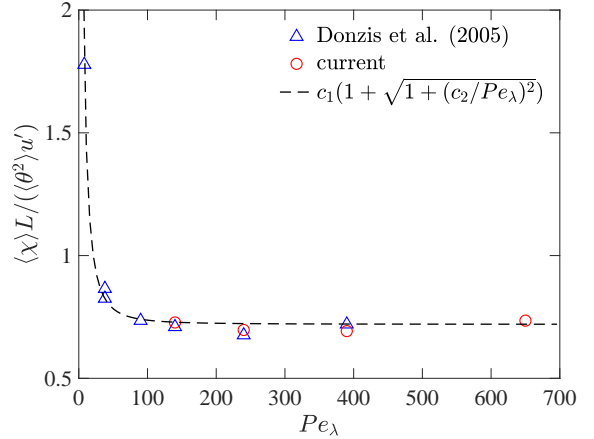}
\caption{
Normalized scalar dissipation rate for $Sc=1$, as a function of microscale
P\'eclet number $Pe_\lambda \propto 1/D$. 
The data in (blue) triangles are from \cite{Donzis05}; the new data are described in Table~\ref{tab:data}.
The functional form of the fit to the data in \cite{Donzis05} 
is shown in the legend, with $c_1=0.36$ and $c_2=31$.  
%The statistical errors are less than the symbol height.
%\{Similar result can be expected when 
%$Sc$ is fixed at another value, but the value of $c_1$
%will also change.}
}
\label{fig:sc1diss}
\end{center}
\end{figure}

Since the passive scalar is advected by the underlying velocity field,
investigating scalar dissipation anomaly in principle 
requires the joint limit of $\nu, D \to 0$. 
Specific practical circumstances on how they approach zero,
motivate two separate scenarios.
In the first scenario, we can take the joint
limit such that $Sc$ is a constant and
thus the Reynolds number increases.
For this case, there is some evidence at $Sc = \mathcal{O}(1)$ 
that the scalar dissipation indeed becomes independent
of $D$ \cite{Donzis05}. 
Figure~\ref{fig:sc1diss} reaffirms this by showing that $\langle \chi\rangle$, non-dimensionalized by the large-scale quantity $\langle\theta^2\rangle u^\prime/L$, asymptotes to a constant for 
large Taylor-scale P\'eclet number $Pe_\lambda  = u^\prime \lambda/D$, 
where $u^\prime$ is the {\em rms} of velocity fluctuations, $\lambda = u'/\sqrt{\langle(\partial u/\partial x)^2\rangle}$ is the Taylor microscale, and $L$ is the integral length-scale.

In the second scenario, either
$\nu$ or $D$ approaches zero faster, such that
the $Sc\to0$ or $\infty$, respectively.
Here, we focus on the latter case of $Sc \to \infty$ 
\footnote{the case of $Sc\to0$ is somewhat 
straightforward and can be expected to 
yield a similar result as $Sc=1$, e.g. see \cite{Donzis05}}.
The mixing of scalars with $Sc\gg1$ is characterized
by the development of very fine scales, even smaller than those
in the velocity field, which are extremely challenging to
resolve in both experiments and simulations \cite{nash02,yeung2004}. 
Consequently, the study of high $Sc$ scalars has been 
historically limited to very low Reynolds numbers, 
where the turbulence is not adequately developed. 
However, even at very low Reynolds numbers, there has been 
some indication that the asymptotically constant values of 
$(\langle \chi \rangle L)/(\langle\theta^2\rangle u^\prime)$ 
become smaller with increasing $Sc$ \cite{Donzis05}.
In this Letter, utilizing new state-of-the-art simulations at significantly
higher Reynolds numbers (corresponding to fully-developed turbulence),
we present new results which demonstrate conclusively that the normalized 
scalar dissipation rate  approaches zero at large $Sc$,
rendering turbulence ineffective at mixing. We additionally show
that this inefficacy is also carried over to the larger scales, with 
important theoretical and practical implications.

%There was already some suggestion in \cite{Donzis05} that the asymptotically constant %values of $(\langle \chi \rangle L)/(\langle\theta^2\rangle u^\prime)$ become smaller %with increasing $Sc$, but the data available at that time were limited to very low %Reynolds number, where the turbulence was
%not adequately developed, and hence inconclusive.

\paragraph{Direct numerical simulations:}
The data utilized here are generated using the canonical DNS setup of isotropic turbulence in a periodic domain \cite{Ishihara09,buaria_nc}, forced at large scales to maintain statistical-stationarity. For the passive scalar, we simultaneously solve the advection-diffusion equation in the presence of uniform mean-gradient $\nabla \Theta = (G,0,0)$ along the Cartesian direction $x$ \cite{PK02}. 
For $Sc=1$, we utilize the conventional Fourier pseudo-spectral methods for both the velocity and scalar fields. For $Sc=4$ and higher, we utilize a hybrid approach \cite{gotoh12a, clay.cpc1, clay.cpc2, clay.omp}, where the velocity field is obtained pseudo-spectrally, focused on resolving the Kolmogorov length scale $\eta$, and the scalar field by using compact finite differences on a finer grid to adequately resolve the smaller Batchelor scale $\eta_B = \eta {Sc}^{-1/2}$. The database is summarized in Table~\ref{tab:data}.
For many cases, we have performed simulations with various
small-scale resolutions to ensure 
accuracy of the statistics  \cite{clay_thesis}.
Our runs also meet the resolution requirements proposed in  
\cite{BPBY2019}. However, we note that 
while \cite{BPBY2019} was focused on studying extreme events,
the statistics reported in this work are not as sensitive 
to resolution \cite{KI18,BBP20}.

\begin{table}
\begin{tabular}{c|c|c|c|c|c|c}
$\re$ & $Sc$  & $N_v^3$   & $k_{max}\eta$ & $N_\theta^3$ & $k_{max}\eta_B$ & $T_{sim}/T_E$ \\
\hline
 140 & 1     & $512^3$   & 3    & $512^3$   & 3    &  10    \\
 140 & 4     & $512^3$   & 3    & $1024^3$  & 3    &  90    \\
 140 & 4     & $512^3$   & 3    & $2048^3$  & 6    &  27    \\
 140 & 8     & $256^3$   & 1.5  & $1024^3$  & 2    &  90    \\
 140 & 8     & $512^3$   & 3    & $1024^3$  & 2    &  85    \\
 140 & 8     & $512^3$   & 3    & $2048^3$  & 4    &  45    \\
 140 & 16    & $256^3$   & 1.5  & $1024^3$  & 1.5  &  98      \\
 140 & 16    & $256^3$   & 1.5  & $2048^3$  & 3    &  44    \\
 140 & 16    & $512^3$   & 3    & $1024^3$  & 1.5  &  84      \\
 140 & 16    & $512^3$   & 3    & $2048^3$  & 3    &  56    \\
 140 & 32    & $512^3$   & 3    & $2048^3$  & 2    &  44    \\
 140 & 32    & $512^3$   & 3    & $2048^3$  & 2    &  19    \\
 140 & 32    & $1024^3$  & 6    & $4096^3$  & 4    &  11    \\
 140 & 64    & $512^3$   & 3    & $2048^3$  & 1.5  &  53      \\
 140 & 64    & $1024^3$  & 6    & $4096^3$  & 3    &  9    \\
 140 & 128   & $512^3$   & 3    & $4096^3$  & 2    &  23    \\
 140 & 256   & $1024^3$  & 6    & $8192^3$  & 3    &  6    \\
 140 & 512   & $1024^3$  & 6    & $8192^3$  & 2    &  9    \\
\hline
240 & 1     & $1024^3$  & 3    & $1024^3$  & 3    & 10 \\
390 & 1     & $2048^3$  & 3    & $2048^3$  & 3    & 10\\
390 & 8     & $2048^3$  & 3    & $8192^3$  & 4    & 6 \\
650 & 1     & $4096^3$  & 3    & $4096^3$  & 3    & 10\\
\hline
\end{tabular}
\caption{
Simulation parameters for the DNS runs used in the current work: the Taylor-scale Reynolds number $\re$, the Schmidt number $Sc$, the number of grid points for the velocity and scalar fields, $N_v^3$ and $N_\theta^3$, the spatial resolution for the velocity and scalar fields, respectively $k_{max}\eta$ and $k_{max}\eta_B$, 
and the simulation length $T_{sim}$ in statistically stationary state
in terms of the large-eddy turnover time $T_E$.
For each case, the domain length is $L_0=2\pi$, and $L \approx L_0/6$.
%We have averaged over at least 30 snapshots for large runs (far more for smaller runs),
%which are roughly equally spaced in statistically stationary state -- hence the results
%presented here have excellent statistical convergence.
}
\label{tab:data}
\end{table}

\begin{figure}
\begin{center}
\includegraphics[width=7.5cm]{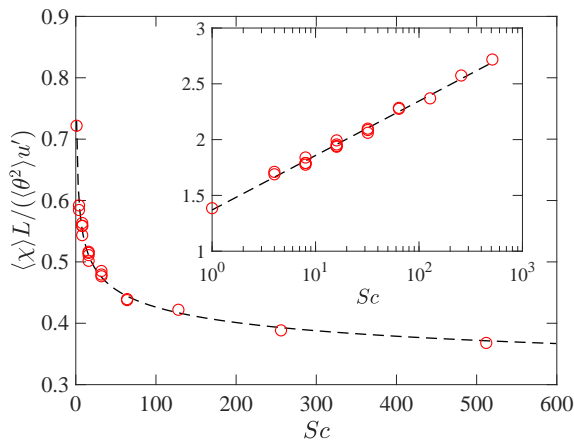}
\caption{
Test for scalar dissipation anomaly at $\re=140$ with increasing $Sc$. The mean scalar dissipation rate is normalized as in Fig.~\ref{fig:sc1diss}. The dashed line corresponds to $1/\log Sc$ dependence. The inset shows the inverse of these data versus $Sc$ on log-linear axes, affirming the $\log Sc$ dependence.
%The statistical errors are less than the symbol height.
}
\label{fig:scdiss}
\end{center}
\end{figure}

\paragraph{Reduction of mixing at diffusive scales:} Here we explore the influence of $Sc$ on mean scalar dissipation rate, $\langle \chi \rangle$. We see in Fig.~\ref{fig:scdiss} that the asymptotic value of scalar dissipation continually decreases with $Sc$. In fact, using arguments based on functional form of the scalar spectrum, the authors of refs.~\cite{BSXDY,Donzis05} showed that the inverse scalar dissipation rate $(\langle\theta^2\rangle u^\prime)/(\langle \chi \rangle L)$ varies as $\log Sc$. In order to see this behavior clearly, we plot the inverse dissipation versus $\log Sc$ in the inset of Fig.~\ref{fig:scdiss}. The data are in excellent agreement with expectations.

The observation that the normalized scalar dissipation tends to zero in the limit $Sc \to \infty$, albeit logarithmically, suggests that the diffusivity is ultimately incapable of smoothing the scalar fluctuations and that there is no mixing at small scales. This picture can be intuitively understood from a Lagrangian perspective by considering trajectories of individual scalar particles \cite{falkovich01,sj2010,SP.2013}. Physically, mixing occurs when some local concentration of scalar particles eventually disperses through the fluid under the combined action of turbulence and molecular diffusion. If we consider two coincident scalar particles, the diffusivity is necessary to create some finite separation, thereafter allowing turbulence to take over; however, in the limit of $D\to0$, they cannot separate and the action of turbulence does not manifest \cite{BSY.2015,BYS.2016}. 
%In fact, the Lagrangian data in \cite{BYS.2016}
%also indicated decreasing scalar dissipation with increasing $Sc$. 

\begin{figure}
\begin{center}
\includegraphics[width=7.2cm]{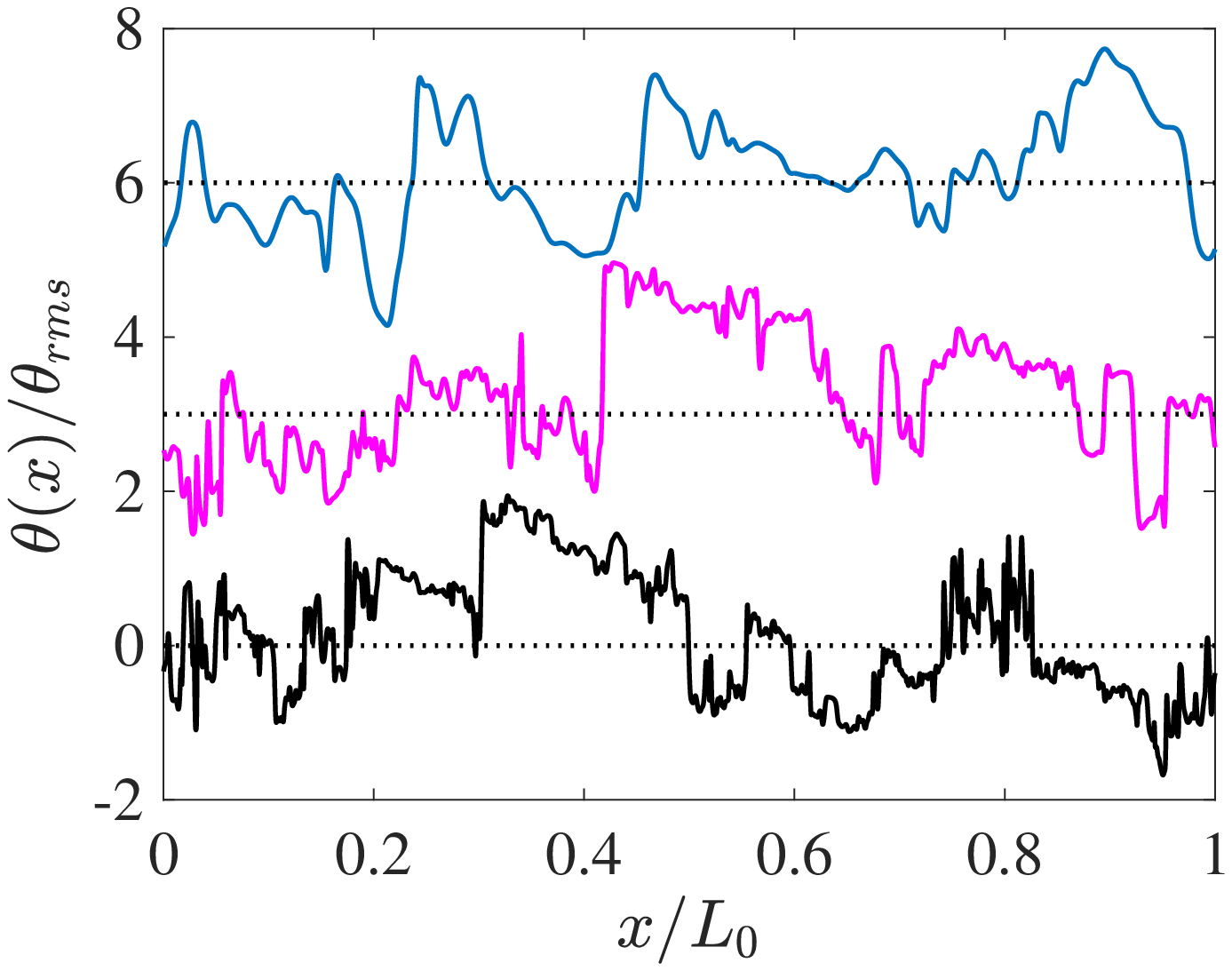} \\
\vspace{0.2cm}
\includegraphics[width=7.2cm]{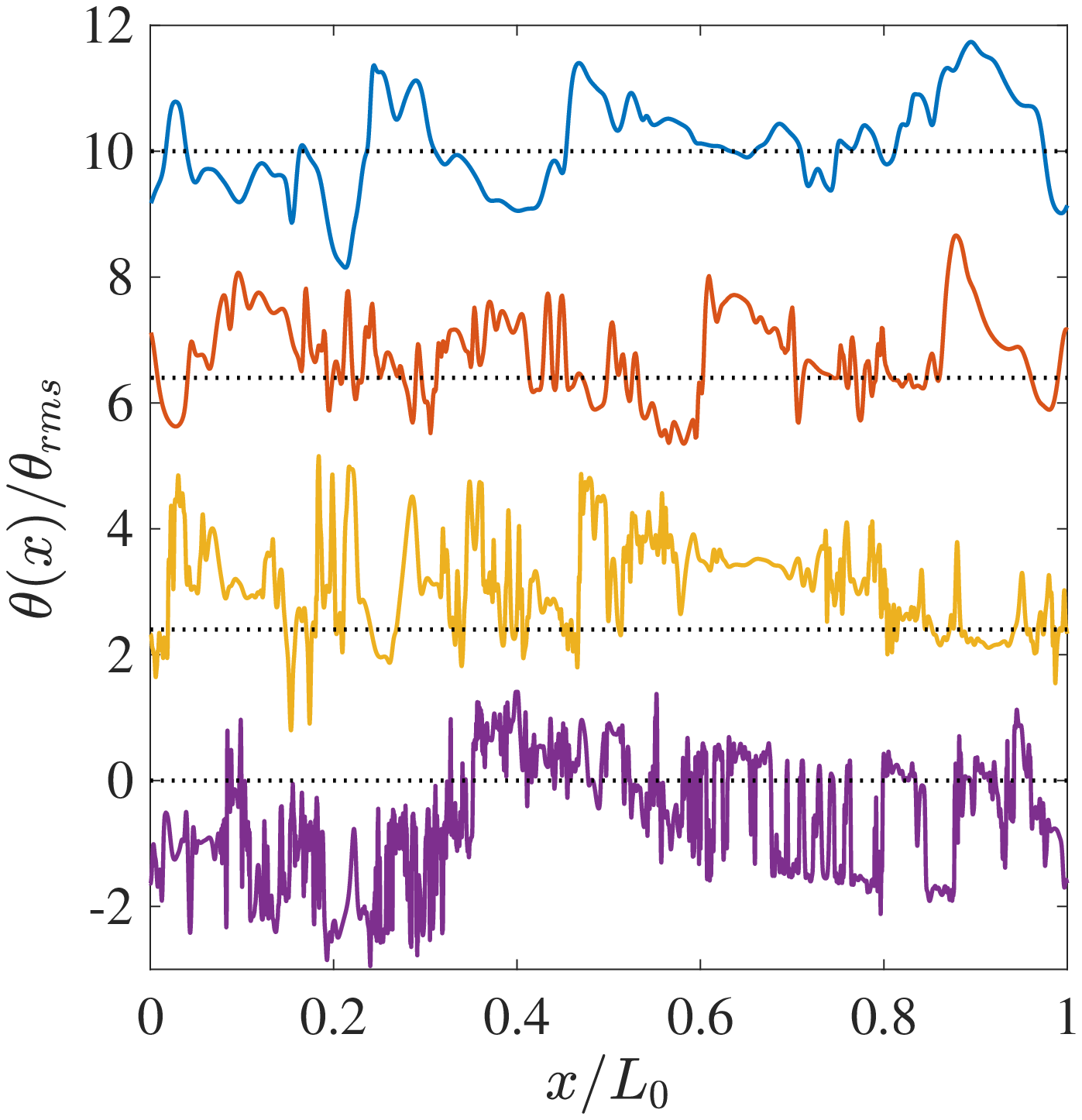}
\caption{
Typical one-dimensional cuts of the scalar field, normalized by the rms, in the direction of the imposed mean gradient ($x$). $L_0 = 2\pi$ is the domain length. The curves in the upper panel correspond to fixed $Sc=1$ and $\re=140$, $390$ and $650$ from top to bottom; those in the lower panel are for fixed $\re=140$ and $Sc=1$, $8$, $64$ and $512$ from top to bottom. The curves are shifted for clarity, as indicated by dotted horizontal lines.
}
\label{fig:rc}
\end{center}
\end{figure}

\paragraph{Reduced mixing at larger scales:} 
Figure~\ref{fig:rc} shows typical 1D cuts of the scalar field in the direction of the mean gradient. The upper panel corresponds to $Sc=1$ and increasing $\re$. The well known ramp-cliff structures (see \cite{KRS91,HS94,CLMV01,sreeni19}) are clearly visible in all traces, with disorganized small-scale fluctuations superimposed on them. With increasing $\re$, small-scale fluctuations expectedly become more conspicuous, but the steep cliffs remain. In the lower panel, the cuts are for $\re=140$ but varying $Sc$. For low to moderate $Sc$, the ramp-cliff structures stand out as before, but the superimposed scalar fluctuations become stronger with increasing $Sc$. The large-scale ramp-cliff structures seemingly continue to be present even at the highest $Sc$ ($=512$), but are overwhelmed by sharp oscillations essentially between the smallest and largest concentrations, leading to inefficient mixing at larger scales.

It is worth noting that 
the scalar dissipation also represents the
scalar variance flux from the large scales
through intermediate (inertial) scales to the smallest
(analogous to the energy dissipation  representing
the flux of kinetic energy). Since the inertial range dynamics
are not influenced by either $\nu$ or $D$, 
in principle the dimensional scalar dissipation 
can still be non-zero as $Sc$ increases. However, in contrast
the scalar variance increases with $Sc$
(ostensibly through a broadening viscous-convective range),
and thus causes the normalized scalar dissipation to approach zero.
In other words, as $Sc$ is increased, turbulence responds not only be
producing strong scalar gradients, but even stronger scalar fluctuations,
which ultimately lead to inefficient mixing.

%Thus, even though the turbulence 
%responds to reduction of scalar diffusivity by producing 
%stronger scalar
%gradients, as anticipated
%from anomalous dissipation, it also generates stronger scalar fluctuations,
%which ultimately lead to reduced mixing. 

%The standard interpretation of energy dissipation is that it is the ultimate %manifestation of energy flux from the large scales through intermediate ranges to the %smallest. A similar interpretation holds for scalar dissipation. The vanishing of the %normalized scalar dissipation rate with increasing $Sc$ not only suggests 
%that mixing is reduced at the smallest scales, but also that there is reduced 
%flux of the scalar variance at larger scales where the advective action 
%of turbulence overwhelms the role of diffusivity. 

\paragraph{Structure functions:} To further analyze the reduction in mixing, 
we consider the scalar increment $\Delta_r \theta$ between two points separated by distance $r$, whose moments are the scalar structure functions. In the so-called inertial-convective range, the $p$-th order structure function is expected to follow a power law of the form $\langle (\Delta_r \theta)^p \rangle \sim r^{\zeta_p}$, where $\zeta_p$ is anomalous with respect to the Kolmogorov phenomenology (i.e., $\zeta_p = p/3$) \cite{MY.II,ZW00,GY.2013}. In order to extract $\zeta_p$, we have followed an analysis similar to the recent work \cite{KI18} where $\zeta_2$ was obtained by a power law fit in the inertial-convective range, and higher order moments were extracted through extended self-similarity
\footnote{Although not shown here, we have performed extensive tests (as done in \cite{KI18}) to establish statistical convergence of structure functions of high orders}.

\begin{figure}
\begin{center}
\includegraphics[width=0.45\textwidth]{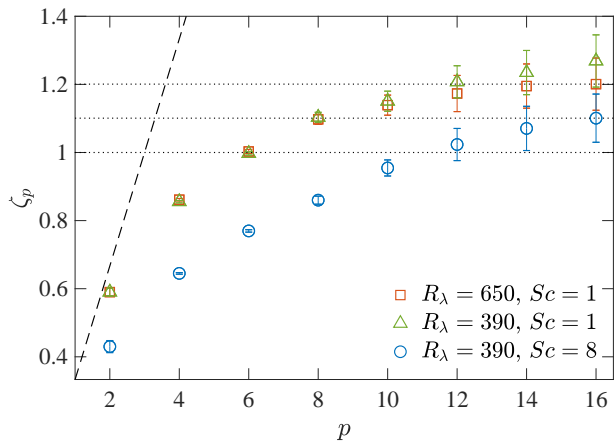}
\caption{
The scalar increment exponent, $\zeta_p$, as a function of the moment order $p$ for 
$\re$ and $Sc$ shown in the legend. The error bars indicate $95\%$ confidence interval. The dotted lines at 1.2 and 1.1 correspond to plausible saturation 
values at $Sc = 1$, $8$, whereas the dotted line at 1 is the likely saturation value 
at $Sc\to\infty$. The dashed line, $\zeta_p=p/3$, 
corresponds to the Kolmogorov phenomenology.
}
\label{fig:exp}
\end{center}
\end{figure}

The scaling exponents $\zeta_p$ are plotted against the moment order in Fig.~\ref{fig:exp}, for $\re \ge 390$. The results for $\re=650$ and $Sc=1$ are virtually identical to those of \cite{KI18}, and reaffirm that the scalar exponents saturate to $\lim_{p\to\infty} \zeta_p  = \zeta_\infty \approx1.2$.  In comparison, the exponents for $\re=390$ and $Sc=1$ are mostly identical to those at $\re=650$, but differ somewhat for $p\ge12$ (possibly due to a slightly smaller scaling range from which the exponents were extracted). The more important result is that for $\re=390$ and $Sc=8$ the exponents are consistently smaller than those for $Sc=1$ and 
tend to saturate at a smaller value of $\zeta_\infty \approx1.1$. Evidently, the smaller saturation value for larger $Sc$ invites the question as to whether it is bounded as $Sc \to \infty$. 

\begin{figure}
\begin{center}
\includegraphics[width=0.42\textwidth]{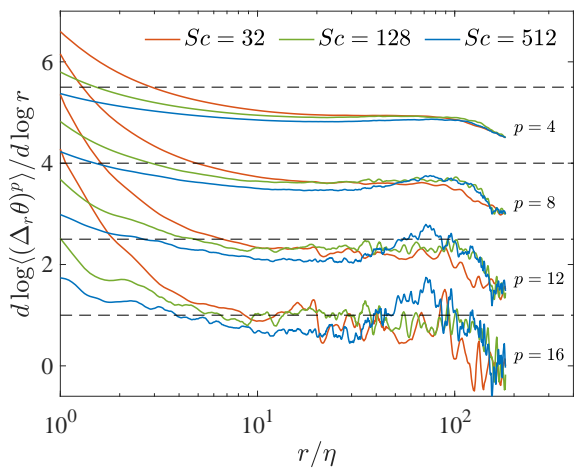}
\caption{
The local slope of $p$-th order scalar structure functions at $\re=140$ and $Sc=32,128$ and $512$. The curves are shown for $p=4,8,12$ and $16$. They are shifted vertically for clarity and the corresponding dashed lines represent a local slope of unity. 
}
\label{fig:ls}
\end{center}
\end{figure}

For a definitive answer, one needs to obtain data for higher $Sc$ for at least $\re = 650$ (at which convincing scaling exists). But large $Sc$ at  $\re=650$ are unlikely to be attainable anytime soon. We have therefore analyzed the data at lower $\re=140$, 
for which inertial range characteristics just begin to manifest \cite{yeung97,Ishihara09}. In Fig.~\ref{fig:ls}, we show the local slope of the 
structure functions for orders $p=$4, 8, 12 and 16 at $Sc$=32, 128 and 512 
(the curves for different $p$ are shifted for clarity).
% and the dashed lines represent local slopes of unity. 
With increasing $p$, the curves for all $Sc$ progressively get closer to local
slope of unity. If we focus on the region $r/\eta \gtrsim 30$, which nominally corresponds to onset of the inertial-convective range \cite{KI18}, it appears that the local slope for all $Sc$ are approximately equal for highest $p$ values, and close to unity---hinting that the high-order exponents saturate at about $1$ as $Sc \to \infty$.

\begin{figure}
\begin{center}
\includegraphics[width=0.45\textwidth]{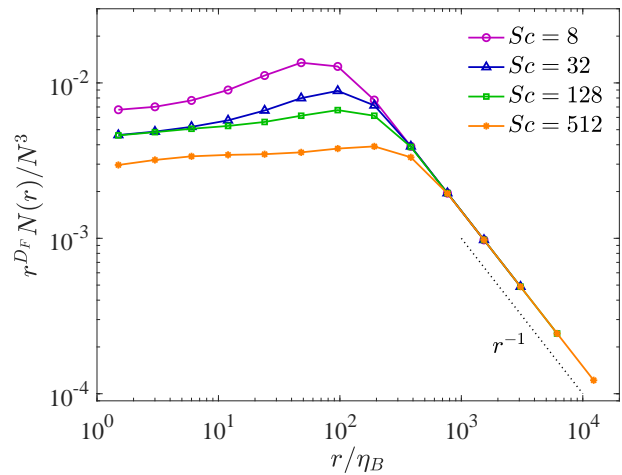}
\caption{
Compensated plot of $N(r)$, the number of cubes of side $r$ containing the scalar fronts satisfying the threshold condition $|\partial \theta/\partial x| \ge 0.2 \theta_{rms}/\eta_B$ \cite{KI18}. Curves are shown for $Sc$=8, 32, 128 and 512 for $\re=140$. $N^3$ is the total number of grid points. $D_F=3-\zeta_\infty$ is the fractal co-dimension. We set $D_F=2$ corresponding to $\zeta_\infty=1$.
%The statistical errors are less than the symbol height.
}
\label{fig:boxc}
\end{center}
\end{figure}

%Remarkably, this gives $\langle (\Delta_r \theta)^p \rangle \sim r^1$, which implies that the scalar increments
%are {\em regular} in the large $Sc$ limit.

\paragraph{Co-dimension result:} Finally, we turn to quantifying the fractal-dimension of sharp scalar fronts and understanding how it relates to the saturation exponent. In \cite{KI18}, the authors found that $\zeta_\infty$ and the box-counting dimension $D_F$ of the sharp scalar fronts (satisfying the threshold $|\partial \theta/\partial x| \ge 0.2 \theta_{rms}/\eta_B$), add up to the Euclidean dimension of the flow, i.e., $\zeta_\infty+D_F = 3$. In that same spirit, we perform box-counting of the strong scalar gradients corresponding to sharp fronts, given by $N(r)$ for various cubes of edge size $r$. For the saturation exponent $\zeta_\infty=1$, the co-dimension corresponds to $D_F=2$. In Fig.~\ref{fig:boxc}, we plot the $N(r)/N^3$ compensated by $r^{D_F}$ with $D_F=2$, for the same cases shown in Fig.~\ref{fig:ls}. Remarkably, the curves at the highest $Sc$ exhibit an extended plateau for small scales, consistent with a fractal-dimension of 2. For large $r$, all curves are consistent with $D_F=3$, as expected by the space filling nature at large scales. This consolidates the result that  fractal-dimension of sharp fronts is the co-dimension of the saturation-exponent of scalar structure functions.

%Remarkably, we find that all the curves exhibit a plateau in the intermediate range of $r$, %
%consolidating the result that the fractal dimension of sharp fronts is the co-dimension of %the saturation exponent of scalar structure functions. 

\paragraph{Conclusions:} 
We have demonstrated by several means that  
fully-developed turbulence, 
which enabled effective mixing at unity $Sc$,  
becomes an ineffective mixer when $Sc$ is large. 
The scalar-dissipation-rate, 
when non-dimensionalized 
by large-scale quantities, decreases
with $Sc$ and the scalar field 
effectively oscillates between the largest and smallest 
concentrations without producing  many intermediate levels.
We find that the exponents of  scalar structure functions 
saturate for high-order moments; the saturation 
value appears to be bounded by unity,
which is also confirmed by
showing that large excursions in $\partial \theta/\partial x$ 
have a co-dimension of 2. These results form an important ingredient 
in a fuller understanding of turbulent mixing, and we note that models like 1D-Burger
s equation \cite{bec07} and Kraichnan's passive scalar \cite{balk98} have the same 
behavior of saturated exponents for large moment orders, leveling off at unity.

From a theoretical perspective, our results invite 
revisions to existing phenomenology of scalar turbulence (for large $Sc$). 
While we have considered mixing of passive scalars, 
it would be instructive to extend these results 
to active scalars at large $Sc$, 
e.g. salinity in the ocean ($Sc\sim700$).
In oceanic mixing, it is often assumed that the turbulent 
flux of salinity is equal to that of heat, despite the latter
occurring at $Sc\sim 7$. However, the current study,
together with the work of 
\cite{nash02}, provides strong evidence against it.
On a related note, 
it has been shown in a subsequent analysis \cite{BCSY20b}
that the results reported 
here are seemingly connected to a $Sc$-correction to the
Batchelor length scale, which can play an important role
for both passive and active scalars.

\begin{acknowledgments}

\paragraph*{Acknowledgments:} 
We thank Kartik Iyer and J\"org Schumacher for useful discussions and Kiran Ravikumar for providing the $Sc = 256$ datapoint used in Fig.~\ref{fig:scdiss}. This research used resources of the Oak Ridge Leadership Computing Facility (OLCF), which is a Department of Energy (DOE) Office of Science user facility supported under Contract DE-AC05-00OR22725. We acknowledge the use of advanced computing resources at the OLCF under 2017 and 2018 INCITE Awards. Parts of the data analyzed in this work were obtained through National Science Foundation (NSF) Grant ACI-1036170, using resources of the Blue Waters sustained petascale computing project, which was supported by the NSF (awards OCI- 725070 and ACI-1238993) and the State of Illinois. DB also gratefully acknowledges the Gauss Centre for Supercomputing e.V. (www.gauss-centre.eu) for providing computing time on the supercomputer JUWELS at J\"ulich Supercomputing Centre, where some of the $Sc=1$ simulations were performed.

\end{acknowledgments}

%\bibliography{large_grad,zebib}

%merlin.mbs apsrev4-1.bst 2010-07-25 4.21a (PWD, AO, DPC) hacked
%Control: key (0)
%Control: author (0) dotless jnrlst
%Control: editor formatted (1) identically to author
%Control: production of article title (0) allowed
%Control: page (1) range
%Control: year (0) verbatim
%Control: production of eprint (0) enabled
%

\end{document}